\documentstyle[11pt,aaspp4]{article}

\def\gsim{\;\rlap{\lower 2.5pt
 \hbox{$\sim$}}\raise 1.5pt\hbox{$>$}\;}
\def\lsim{\;\rlap{\lower 2.5pt
   \hbox{$\sim$}}\raise 1.5pt\hbox{$<$}\;}
\newcommand\beq{\begin{equation}}
\newcommand\eeq{\end{equation}}

\begin{document}

\title{X-Ray Absorption by the Hot Intergalactic Medium}
\author{Rosalba Perna and Abraham Loeb}
\medskip
\affil{Harvard-Smithsonian Center for Astrophysics, 60 Garden Street,
Cambridge, MA 02138}

\begin{abstract}
The current census of observed baryons in the local Universe is still
missing a significant fraction of them according to standard Big-Bang
nucleosynthesis. Numerical simulations predict that most of the missing
baryons are in a hot intergalactic medium, which is difficult to observe
through its X-ray emission or Sunyaev-Zel'dovich effect.  We show that the
next generation of X-ray satellites will be able to detect this gas through
the X-ray absorption lines imprinted by its highly-ionized metals on the
spectrum of a background quasar. For the metallicity typically found in
intracluster gas, up to 70\% of the baryons produce O VIII absorption lines
with an equivalent width $\ga 0.1$ eV. The spectrum of any high redshift
quasar is expected to show several such lines per unit redshift due to
intervening gaseous halos of galaxy groups. These lines will be detectable
at a signal-to-noise ratio $\ga 5$ after a day of integration with the
future Constellation-X telescope for any of the several hundred brightest
quasars across the sky.

\end{abstract}

\section{Introduction}

The best estimate for the sum of the baryonic mass currently observed in
stars, neutral hydrogen, and X-ray emitting gas in clusters of galaxies
(Fukugita, Hogan, \& Peebles 1998) falls short of the most-likely value
predicted by standard Big-Bang nucleosynthesis (Burles \& Tytler 1998).
Numerical simulations suggest that most of the missing baryons reside in a
hot intergalactic medium, which is too rarefied to be detectable through
its X-ray emission or the Sunyaev-Zel'dovich effect (e.g., Ostriker \& Cen
1996; Theuns et al.  1998).  Much of this gas resides in the outskirts of
galaxy groups or clusters, and hence is likely to possess a similar
metallicity to that observed in the X-ray emission spectrum from the cores
of these systems (Mushotzky et al.  1996).  In this paper we explore the
possibility of detecting this hot intergalactic medium through the X-ray
absorption lines imprinted by its metals on the spectrum of a background
quasar.

Detection of X-ray absorption lines from galaxy clusters had been suggested
in the past as a means of measuring distances (Krolik \& Raymond 1988;
Sarazin 1989).  However, because of the limited sensitivity of X-ray
instruments at that time, these early papers restricted their attention to
the strong lines produced in the cores of rich clusters and hence to the
unlikely superposition of a quasar behind such cores. The next generation
of X-ray telescopes, manifested at their best by the successor to AXAF,
Constellation-X (see http:// constellation.gsfc.nasa.gov), will achieve a
much greater sensitivity than previously imagined. It is therefore the
purpose of this paper to quantify the full probability distribution of
equivalent widths for the intergalactic X-ray absorption lines imprinted on
the spectrum of any {\it random} quasar.  Since much of the baryonic mass
is currently assembled into virialized objects and since the present
nonlinear mass-scale is that of galaxy groups, most of the absorption lines
would be produced by gas in the outskirts of these groups.  In \S 2 we
combine the Press-Schechter mass function and the
singular-isothermal-sphere model to predict the density and temperature
distribution of this gas in a standard CDM cosmology with
$\sigma_{8h^{-1}}=0.5$.  We used a Bardeen et al.  (1986) power spectrum
with a tilted primordial index of $n=0.7$. Because the power-spectrum is
normalized on the cluster mass scale, our results depend only weakly on the
choice of the power spectrum shape (and agree by construction with X-ray
data for the temperature function of groups and clusters; see, e.g. Henry
et al.  1995).  We assume an $\Omega=1$, $\Omega_b=0.1$, $h=0.6$ cosmology
throughout the discussion.  In \S 3 we compute the corresponding statistics
of the absorption lines produced by highly-ionized metals in this gas.
Finally,
\S 4 summarizes the main conclusions of this work.

\section{Statistical Properties of the Gas in the 
Outskirts of Groups and Clusters of Galaxies}

Observations of the X-ray surface brightness of galaxy clusters are
commonly fitted by the isothermal $\beta$-model. This model assumes that
the gas temperature is constant and the radial profile of its mass density
is given by
\begin{equation}
\rho_{\rm gas}(r,T)
=\rho_{0}(T)\left[1+\left(\frac{r}{r_c}\right)^2\right]^{-3\beta/2}\;,
\label{eq:ro}
\end{equation}
with $r_c$ being the core radius. Typical values for the parameters of this
functional fit are clustered around $r_c \approx 0.25$ Mpc and
$\beta\approx 2/3$ (Jones \& Forman 1984); for specificity, we adopt these
values throughout our discussion.  The central density can be matched at
large radii far from the core ($r\gg r_c$), to that of a singular
isothermal profile, $\rho_{\rm gas}\approx f_{\rm gas} kT/(2\pi \mu m_p G
r^2)$, where $f_{\rm gas}=\Omega_b/\Omega$ is the global mass fraction of
the gas (assumed to equal the universal baryon-to-total matter density
ratio), $T$ is the gas temperature,
$m_p$ is the proton mass, and  
$\mu\approx 1.22$ is the mean atomic weight. Hence we set, 
\begin{equation}
\rho_{0}(T)=\frac{f_{\rm gas}kT}{2\pi\mu m_pGr^2_c}\;.
\label{eq:ro0}
\end{equation}
We implicitly assume that the dark matter provides the gravitational
potential needed to support the above gas distribution, $\phi= -(kT/\mu
m_p)\ln \rho_{\rm gas}+ {\rm const}$.

At any given redshift, the comoving number density of galaxy groups or
clusters with a temperature between $T$ and $T+dT$, namely
$[dn(T,z)/dT]dT$, can be derived by combining the Press-Schechter (1974)
mass function, $dn(M,z)/dM$, with the total mass-temperature relation
calibrated from numerical simulations (e.g., Pen 1998),
\begin{equation}
M(T,z)=M_8\left(\frac{T}{T_8(1+z)}\right)^{3/2}\;,
\label{eq:M-T}
\end{equation}
with $kT_8=4.9$ keV (for $\Omega=1$) and $M_8=4\pi\rho_{\rm crit}
(8h^{-1}{\rm Mpc})^3/3$ is the mass inside a sphere of radius $8h^{-1}$ Mpc
filled with the cosmological critical density $\rho_{\rm crit}$.  Each
cluster is assigned a cut-off radius, $R_{\rm cut-off}$, which defines its
total mass.  Typically $R_{\rm cut-off}\gg r_c$, and so $M(T,z)\approx
2kTR_{\rm cut-off}/(\mu m_p G)$.

The column density of gas particles along the line of sight to a background
source which is projected at an impact parameter $b$ relative to the
cluster center is given by
\begin{equation}
N_{\rm gas}(b,T)= N_0(T)\left[1+\left(b/r_c\right)^2\right]^{-1/2}\;,
\label{eq:Ngas}
\end{equation}
with $N_0(T)\equiv \pi \rho_{0}(T) r_c/\mu m_p$.  We denote by $Z(X)\equiv
N_X/N_{\rm H}$ the fractional abundance of a heavy element $X$ relative to
hydrogen, and by $\Upsilon \equiv N^i/N_X$ the fraction of this element
which is ionized $i$ times. In collisional equilibrium, the fractional
abundance of the ion $X^i$ is only a function of temperature (Sarazin \&
Bahcall 1977), i.e. $\Upsilon=\Upsilon (T)$. We ignore photoionization by
the X-ray background since we focus on the high density environments of
virialized objects (in which the highest equivalent-width lines are
expected to occur) where collisional ionization dominates. The collisional
ionization/recombination time is typically much shorter than the Hubble
time at the temperatures ($\ga 0.5$ keV) and overdensities ($\ga 10$) of
interest here.
Assuming a uniform
metallicity, the column density of the ion $X^i$ is then
\begin{equation}
N^i(b,T)=0.92 \;N_{\rm gas}(b,T)\,Z(X)\,\Upsilon(T)\;,
\label{eq:Ni}
\end{equation}  
where 0.92 is the 
fraction of hydrogen atoms by number.

The equivalent width of the absorption line produced by an ion
column-density $N^i$ is given by
\begin{equation}
W^i=E \frac{\Delta v_X}{c}\,g(\tau)\;,
\label{eq:W}
\end{equation}
where 
\[g(\tau)\equiv\int_{-\frac{c}{\Delta v_X}}^\infty
dy \left[1-\exp\left(-\tau \phi(y)\right)\right]\,,
\;\;\;\;\;\;\;\; \phi(y)=\frac{1}{\sqrt{\pi}}\exp(-y^2)\]
and
\[\tau\equiv N^i\frac{2\pi^2\alpha\lambda_c^2 f_i}
{(E/m_ec^2)(\Delta v_X/c)}\;.\] Here $E$ is the transition energy, $f_i$ is
the oscillator strength for absorption, $m_e$ is the electron rest mass,
$\lambda_c=3.86\times 10^{-11}$ cm is the electron Compton wavelength, and
$\Delta v_X$ is the velocity dispersion of the element $X$. For a thermally
broadened line $\Delta v_X={\sqrt{2kT/A(X) m_p}}$ (Rybicki \& Lightman
1979), where $A(X)$ is the atomic number of the element $X$.

\section{Statistics of X-Ray Absorption Lines}

Next we would like to find the mass fraction of the hot intergalactic gas
that can be probed by an X-ray telescope of a given sensitivity.  We define
$W_{\rm min}$ to be the minimum equivalent width of an absorption line that
can be resolved by the instrument.  For a cluster temperature $T$, the
column density $N^i$ of a given ion that would yield absorption
with an equivalent width $W_{\rm min}$ is found by inverting
equation~(\ref{eq:W}).  The impact parameter corresponding to this density
can be found from equations (\ref{eq:Ngas}) and (\ref{eq:Ni}), namely
$b(W_{\rm min},T)=r_c\{\left[ {N_0^i(T)}/{N^i(W_{\rm min},T)}\right]^2
-1\}^{1/2}$, where $N_0^i(T)\equiv N^i(b=0,T)$.  Finally, the total gas
mass enclosed within the cylinder defined by this impact parameter is given
by
\begin{equation}
M_{\rm gas}(>W_{\rm min},T)=2\pi^2\rho_0(T)r^3_c\left\{
\left[1+\left(\frac{b(W_{\rm min},T)}{r_c}\right)^2\right]^{1/2}-1\right\}\;. 
\label{mass}
\end{equation}
The fraction of the total baryonic mass 
that is traced at a redshift $z$ through X-ray observations with an
equivalent-width sensitivity $W_{\rm min}$ is then
\begin{equation}
\frac{\Omega_{\rm gas}(>W_{\rm min},z)}{\Omega_{\rm b}}=
\frac{1}{\Omega_{\rm b}\rho_{\rm crit}}\int\;dT\frac{dn(T,z)}{dT}
M_{\rm gas}(>W_{\rm min},T)\;.
\label{eq:om}
\end{equation}

Figure 1 shows the baryonic mass fraction expressed in equation
(\ref{eq:om}) at two redshifts, namely $z=0$ in panel (a) and $z=1$ in
panel (b).  In both panels, the solid curve shows the result for the
$\lambda=18.97$\AA~absorption line of O VIII and the dashed curve
refers to the $\lambda=1.85$\AA~transition of Fe XXV. In this
calculation we have assumed that the spectral lines are only thermally
broadened, since the lines are generically not saturated and so turbulent
broadening has a negligible influence on their equivalent width. The
fractional abundance $\Upsilon(T)$ has been computed by interpolating the
data from the tables of Arnaud \& Rothenflug (1985) for O VIII and of
Arnaud \& Raymond (1992) for Fe XXV. We have assumed the metallicity of the
gas to be $0.3 Z_\odot$ for iron and $0.5Z_\odot$ for oxygen (see, e.g.
Mushotzky et al. 1996; Gibson et al. 1997; Fukuzawa et al.  1998) with the
solar metallicity abundances taken from Anders \& Grevesse (1989).

Figure 1a shows that at low redshifts most of the collapsed baryons can be
probed by an instrument with a sensitivity $\ga 0.1$ eV to the O VIII line.
The fraction is lower but still substantial at moderate reshifts, as shown
in Figure 1b.  Overall, the baryonic mass probed in absorption extends up
to an order of magnitude above the value probed by the conventional methods
of detecting X-ray emission (Fukugita et al. 1997).  The O VIII line traces
more baryons than the Fe XXV line, since the oxygen ion exists in the
temperature regime of $0.1$--$1$ keV while the iron ion is abundant only at
temperatures $\ga 1$ keV (Arnaud \& Rothenflug 1985).  The lower
temperature regime is characteristic of galaxy groups (Henry et al.  1995;
Burns et al.  1996), which contain most of the mass in the present-day
Universe.

In order to assess the feasibility of these observations it is useful to
estimate the expected number of absorbers of a given equivalent width per
unit redshift. The differential cross-section for intercepting an ion
column density between $N^i$ and $N^i + dN^i$ within a given cluster of
temperature $T$ is given by $[d\sigma/dN^i](N^i)=2\pi
b(N^i)|db/dN^i|_{b=b(N^i)}$.
Using equations~(\ref{eq:Ngas}) and~(\ref{eq:Ni}) we get
\begin{equation}
{d\sigma\over d N^i}(N^i,T) = \left\{
  \begin{array}{ll}
  2\pi r^2_c {[N^i_0(T)]^2}/{(N^i)^3},  
& \hbox{$N^i_{\rm min}(T)\le N^i\le N_0^i(T)$} \\ 
      0 \,, & \hbox{otherwise.} \\
\end{array}\right.
\label{eq:PN}
\end{equation}
where $N^i_{\rm min}(T)$ is the ion column density at $b=R_{\rm cut-off}$.
Correspondingly, the differential cross-section for detecting a resonant
absorption line with an equivalent width between $W^i$ and $W^i+dW^i$ is
given by $[d\sigma/d W^i](W^i)=[d\sigma/d
N^i](N^i(W^i))|dN^i/dW^i|_{N^i=N^i(W^i)}$.  Equation (\ref{eq:W}) then
yields,
\begin{equation}
{d\sigma\over d W^i}(W^i,T) = 
r^2_c\frac{[N^i_0(T)]^2}{[N^i(W^i,T)]^3}
\frac{1}{\pi\alpha m_ec^2\lambda_c^2f_i}
\left[\frac{dg(\tau)}{d\tau}\right]^{-1}_{\tau=\tau(W^i)}\;, 
\label{eq:PW}
\end{equation}
for ${W^i(N^i_{\rm min})\le W^i \le W^i(N^i_0)}$.  Finally, by integrating
over the cluster temperature distribution, we derive the probability of
observing an absorption line with an equivalent width between $W^i$ and
$W^i+dW^i$ per unit redshift at a redshift $z$,
\begin{equation}
{dP\over d W^i dz}(W^i,z) \equiv \int\; dT\; {d\sigma\over d
W^i}(W^i,T)\frac{dn}{dT}(T,z)\left|c\frac{dt}{dz}\right|(1+z)^3\;,
\label{eq:fW}
\end{equation}
where in the adopted $\Omega=1$ cosmology the redshift derivative of cosmic
path length is given by $\vert c dt/dz\vert= cH_0^{-1}(1+z)^{-5/2}$, with
$H_0$ being the Hubble constant.  Figure 2 shows this probability
distribution at the redshifts $z=0$ [panel (a)] and $z=1$ [panel (b)] for
the previously mentioned absorption lines of O VIII (solid line) and Fe XXV
(dashed line).  We find that the spectrum of any {\it random} quasar at
$z=2$ is expected to show several O VIII lines with an equivalent width
$\ga 0.1$ eV due to intervening gaseous halos of galaxy groups.

Finally, we would like to estimate the detectability of the predicted
absorption signal for an X-ray telescope of a given area and spectral
resolution. The maximum signal-to-noise ratio, S/N, is dictated by photon
counting statistics. Given the characteristic surface brightness
distribution of clusters, emission by the intracluster gas can be ignored
at several core radii for a characteristic telescope aperture $\la
1^\prime$.  The required number of continuum photons from the quasar which
the detector must receive within its spectral resolution bin of width
$\Delta E$ around the line is (Sarazin 1989),
\begin{equation}
(\#~{\rm of~photons})~\ga\left(\frac{\rm S}{\rm
N}\right)^2\left(\frac{\Delta E}{W_{\rm min}}\right)^{2}\;.
\label{eq:Ncont}
\end{equation}
Correspondingly, the flux from the quasar at the line energy $E$ needs to
be
\begin{equation}
F_E\ga\frac{E({\rm S/N})^2\Delta E}{A\,Q\,W_{\rm min}^2 t}\;,
\label{eq:FE}
\end{equation}
where $A$ is the collecting aperture of the telescope, $Q$ is the quantum
efficiency of the detector, and $t$ is the integration time of the
observation.  The future Constellation-X telescope with its planned
effective area of $QA\approx 15,000\;{\rm cm}^2$ and resolution of $\Delta
E=1~{\rm eV}$ could therefore detect an O VIII line with $E=0.654/(1+z)$
keV and an equivalent-width $W_{\rm min}=0.1$ eV at a signal-to-noise ratio
$S/N=5$ after $t=10^5$ seconds of integration on a quasar with $F_E\ga
2\times 10^{-12}~{\rm erg~cm^{-2}~s^{-1}~keV^{-1}}$.  Based on the number
counts of quasars brighter than a flux $S$ in the 0.5--3.5 keV band,
$N(>S)\approx 20\times (S/10^{-11}~{\rm erg~cm^{-2}~s^{-1}})^{-1.5} {\rm
sr}^{-1}$ (Maccacaro et al. 1982) and the intrinsic soft X-ray spectrum of
quasars $F_{E}\propto E^{-1.6}$ (Laor et al. 1997), we estimate that there
should be several hundred quasars above the necessary flux threshold across
the sky.  Figure 1 therefore implies that Constellation-X will be able to
probe up to 70\% of the hot gas in the outskirts of groups or clusters of
galaxies through the O VIII resonant absorption line.

\section{Conclusions}

We have investigated the possibility of using the X-ray absorption spectrum
of a background quasar to probe the presence of heavy elements in the outer
regions of groups and clusters of galaxies.  Figure 1 demonstrates that
observations of the O VIII line with an equivalent width sensitivity $\ga
0.1$ eV should be able to probe most of the baryonic material out to the
virial radius of these systems, well beyond the regions where X-ray
emission is currently detected.  We estimate that there should be several
such O VIII absorption lines per unit redshift along the line of sight to
any quasar (cf. Fig. 2).  The future Constellation-X telescope will be able
to detect these lines with a signal-to-noise ratio $\ga 5$ in a day of
integration for any of the few hundred brightest quasars accross the sky.

Our model assumed that the gas in groups or clusters of galaxies is
isothermal. The modest radial decline which was observed recently with ASCA
for the temperature of the gas in the outer parts of clusters (Markevitch
et al.  1998), might make absorption features by ionic species such as O
VII more prominent than O VIII in the X-ray absorption spectra of quasars.
On the other hand, energy input by the same supernovae that enrich the gas
with metals might increase the temperature of X-ray halos beyond the value
adopted in equation~(\ref{eq:M-T}) (which was calibrated based on numerical
simulations of gravitational dynamics only).

In our calculations we have assumed a constant metallicity of the gas; a
reduced metal abundance due to inefficient mixing in the outer parts of
groups and clusters would obviously lower our predicted signal. If
absorption will be detected for some quasars, it would be interesting to
correlate the absorption line redshift and sky position with X-ray or
optical catalogs of galaxy clusters. If a group or a cluster are identified
at the absorption redshift, it would be possible to infer the impact
paramter $b$, and the metallicity $Z$, using equations~(\ref{eq:Ni})
and~(\ref{eq:W}).  Observations of the O VIII line around X-ray clusters at
different redshifts could trace the distribution of metals in the
intergalactic medium and possibly shed light on the epoch and nature of the
metal enrichment process.

\acknowledgements

We thank Richard Mushotzky and George Rybicki for useful discussions. This
work was supported in part by the NASA grant NAG 5-7039 and by a fellowship
from the university of Salerno, Italy.

\begin{figure}[t]
\centerline{\epsfysize=5.7in\epsffile{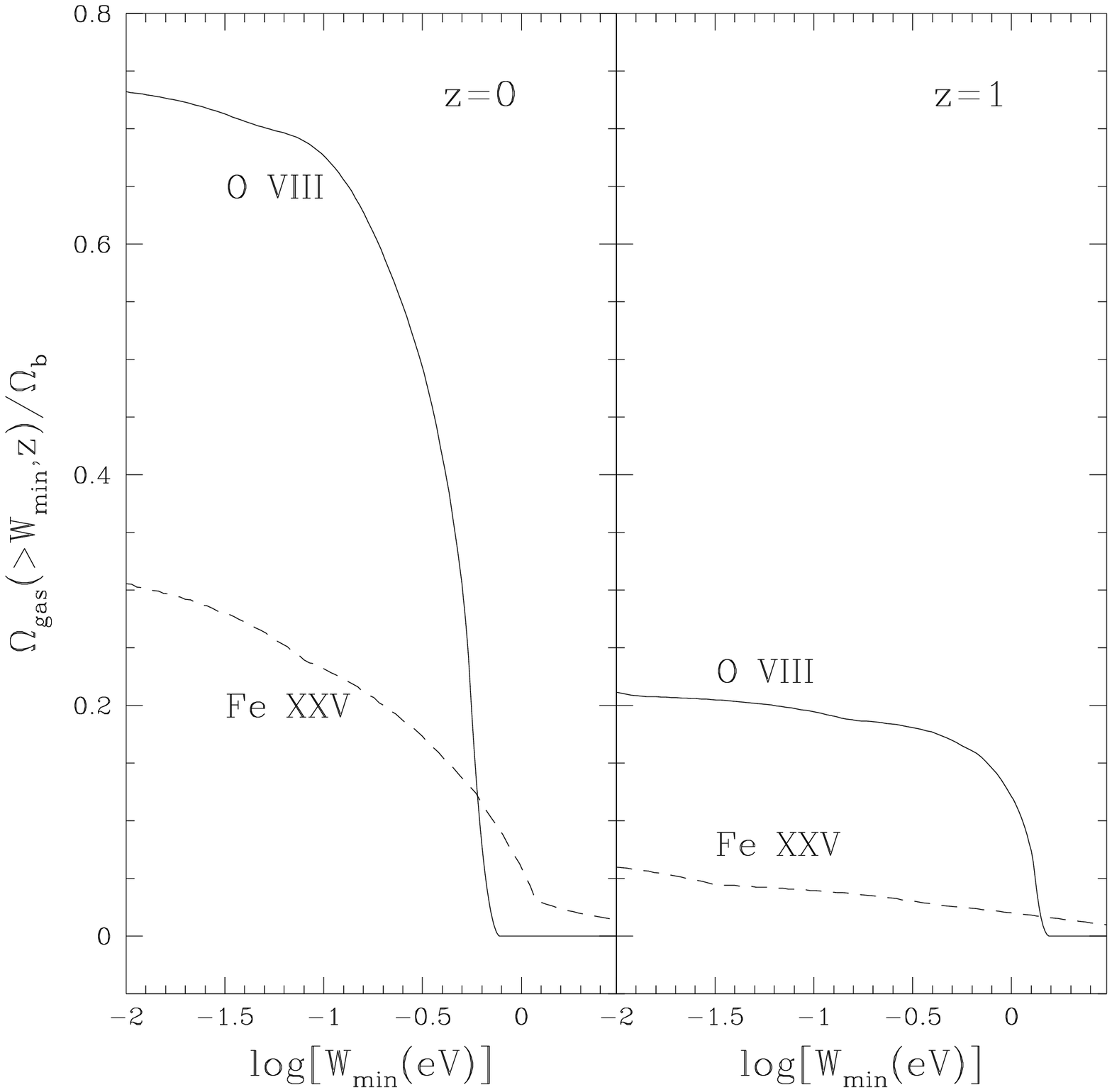}}
\caption{The fraction of baryons in groups or clusters of galaxies that 
can be probed by an X-ray telescope with an equivalent-width sensitivity
$W_{\rm min}$ to absorption lines. The solid curve refers to absorption by
the strongest resonant transition of O VIII [at $0.654/(1+z)$ keV, where $z$
is the absorption redshift], while the dashed curve corresponds to the
strongest resonant absorption by Fe XXV [at $6.7/(1+z)$ keV].}
\label{fig:1}
\end{figure}

\begin{figure}[t]
\centerline{\epsfysize=5.7in\epsffile{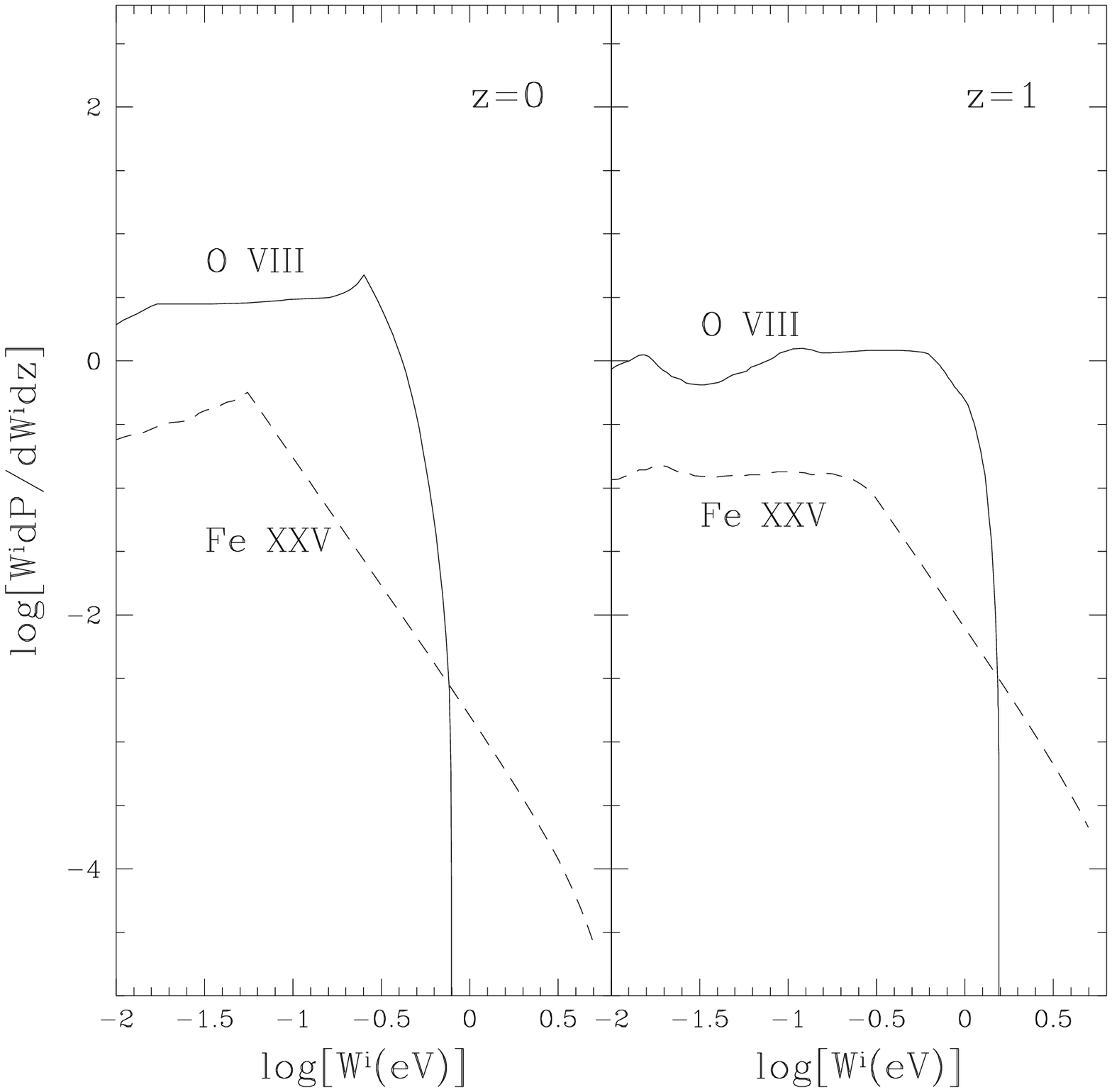}}
\caption{The probability for observing an absorber equivalent width
$W^i$ per unit redshift per logarithmic equivalent-width interval as a
function of the equivalent width. }
\label{fig:2}
\end{figure}

\end{document}